\documentclass{elsart}
\usepackage{epsfig}
\usepackage{amssymb}
\usepackage{rotating}

\begin{document}

\begin{frontmatter}

\title{The ZEPLIN-III dark matter detector: instrument design, manufacture and commissioning}

\author[ITP]{D. Yu. Akimov},
\author[RAL]{G. J. Alner},
\author[ICL,RAL]{H. M. Ara\'ujo},
\author[ICL]{A. Bewick},
\author[ICL,RAL]{C. Bungau},
\author[ITP]{A. A. Burenkov},
\author[SHE]{M. J. Carson},
\author[LIP]{V. Chepel},
\author[UCL]{D. Cline},
\author[ICL]{D. Davidge},
\author[SHE]{J. C. Davies},
\author[SHE]{E. Daw},
\author[ICL]{J. Dawson},
\author[RAL]{T. Durkin},
\author[ICL,RAL]{B. Edwards},
\author[SHE]{T. Gamble},
\author[EDI]{C. Ghag},
\author[SHE]{R. J. Hollingworth},
\author[ICL]{A. S. Howard},
\author[ICL]{W. G. Jones},
\author[ICL]{M. Joshi},
\author[SHE]{J. Kirkpatrick},
\author[ITP]{A. Kovalenko},
\author[SHE]{V. A. Kudryavtsev},
\author[ITP]{I. S. Kuznetsov},
\author[SHE]{T. Lawson},
\author[ICL]{V. N. Lebedenko},
\author[RAL]{J. D. Lewin},
\author[SHE]{P. Lightfoot},
\author[LIP]{A. Lindote},
\author[ICL]{I. Liubarsky},
\author[LIP]{M. I. Lopes},
\author[RAL]{R. L\"{u}scher},
\author[SHE]{J. E. McMillan},
\author[SHE]{B. Morgan},
\author[SHE]{D. Muna},
\author[EDI]{A. S. Murphy},
\author[LIP]{F. Neves},
\author[SHE]{G. G. Nicklin},
\author[SHE]{S. M. Paling},
\author[SHE]{D. Muna},
\author[LIP]{J. Pinto da Cunha},
\author[EDI]{S. J. S. Plank},
\author[RAL]{R. Preece},
\author[ICL]{J. J. Quenby},
\author[SHE]{M. Robinson},
\author[LIP]{C. Silva},
\author[LIP]{V. N. Solovov},
\author[RAL]{N. J. T. Smith},
\author[RAL]{P. F. Smith},
\author[SHE]{N. J. C. Spooner},
\author[ITP]{V. Stekhanov},
\author[ICL]{T. J. Sumner\corauthref{cor1}},
\corauth[cor1]{Corresponding author; address: Astrophysics Group,
Blackett Laboratory, Imperial College London, SW7 2BW, UK}
\ead{t.sumner@imperial.ac.uk}
\author[ICL]{C. Thorne},
\author[SHE]{D. R. Tovey},
\author[SHE]{E. Tziaferi},
\author[ICL]{R. J. Walker},
\author[UCL]{H. Wang},
\author[TEX]{J. White} \&
\author[ROC]{F. Wolfs}
\address[ITP] {Institute for Theoretical and Experimental Physics, Moscow, Russia}
\address[RAL] {Particle Physics Department, Rutherford Appleton Laboratory, Chilton, UK}
\address[ICL] {Blackett Laboratory, Imperial College London, UK}
\address[SHE] {Physics and Astronomy Department, University of
Sheffield, UK}
\address[LIP] {LIP--Coimbra \& Department of Physics of the University of Coimbra, Portugal}
\address[UCL] {Department of Physics \& Astronomy, University of California, Los Angeles, USA}
\address[EDI] {School of Physics, University of Edinburgh, UK}
\address[TEX] {Texas A\&M University, USA}
\address[ROC] {University of Rochester, New York, USA}

\newpage
\begin{abstract}
We present details of the technical design and manufacture of the
ZEPLIN-III dark matter experiment. ZEPLIN-III is a two-phase xenon
detector which measures both the scintillation light and the
ionisation charge generated in the liquid by interacting particles
and radiation. The instrument design is driven by both the physics
requirements and by the technology requirements surrounding the
use of liquid xenon.  These include considerations of key
performance parameters, such as the efficiency of scintillation
light collection, restrictions placed on the use of materials to
control the inherent radioactivity levels, attainment of high
vacuum levels and chemical contamination control.  The successful
solution has involved a number of novel design and manufacturing
features which will be of specific use to future generations of
direct dark matter search experiments as they struggle with
similar and progressively more demanding requirements.
\end{abstract}


\end{frontmatter}

% Main text %%%%%%%%%%%%%%%
%%%%%%%%%%%%%%%%%%%%%%%%%%%%%%%%%%%%%%%%%%%%

%%%%%%%%%%%%%%%%%%%%%%%%%%%%%%%%%%%%%%%%%%%%%%%%%%%%%%%%%%%%%%%%%%%%%%%
\section{Introduction}

ZEPLIN-III is a two-phase (liquid/gas) xenon detector developed
and built by the ZEPLIN Collaboration,\footnote{Edinburgh
University, Imperial College London, ITEP-Moscow, LIP-Coimbra,
Rochester University, CCLRC Rutherford Appleton Laboratory,
Sheffield University, Texas A\&M, UCLA.} which will try to
identify and measure galactic dark matter in the form of Weakly
Interacting Massive Particles, or WIMPs \cite{sumner05,araujo05a}.
Upon completion of physics testing now underway at Imperial
College, the system will join the ZEPLIN-II \cite{zeplin2} and
DRIFT-IIa \cite{drift2} experiments already operating 1100~m
underground in our laboratory at the Boulby mine (North Yorkshire,
UK).

Two-phase emission detectors based on the noble gases date back
several decades \cite{dolgoshein70}. In last decade, this
technology has gained a new momentum in view of increasing
interest for searching rare events, WIMPs in particular, requiring
both large detection masses and high discrimination against
background. In its previous work, the ZEPLIN Collaboration has
explored the potential of high-field xenon systems to enhance
sensitivity and background discrimination
\cite{sumner99,howard01,akimov03}. The operating principle relies
on different particle species generating different amounts of
vacuum ultra-violet (VUV) scintillation light and ionisation
charge in liquid xenon (LXe). The ratio between these two signal
channels provides a powerful technique to discriminate between
electron and nuclear recoil interactions. WIMPs are expected to
scatter elastically off Xe atoms, much like neutrons, and the
recoiling nucleus will produce a different signature to
$\gamma$-ray interactions and other sources of electron recoils.

WIMP detectors differ from more traditional detectors of nuclear
radiation in that they require: i) extremely low radioactive and
cosmic-ray backgrounds, addressed by the use of radio-pure
materials and operation deep underground; ii) excellent
discrimination of the remaining background events, especially for
electron recoils; iii) a low energy threshold for nuclear recoils,
since the kinematics of WIMP-nucleus scattering results in a very
soft recoil spectrum ($\lesssim$100~keV).

Monte Carlo simulations \cite{davidge03,dawson03} were essential
in key areas to inform the design of the instrument. Acceptable
levels of trace contamination must be set for all detector
materials, requiring simulations of internal and external
backgrounds expected from each component. Cosmic-ray-induced
backgrounds also need careful calculation, since experimental
measurements would require nothing short of a dedicated WIMP
detector. These simulations establish the residual electron/photon
and neutron event rates and spectra. Detailed detector simulations
leading to predicted data timelines can be used to find the level
of discrimination and energy threshold which can realistically be
achieved.  Feedback from this process into the design process has
been essential for ZEPLIN III. In addition, the data produced by
two-phase detectors are often complex, and particular simulations
are required to help extract actual physics parameters. Finally,
realistic datasets help with planning the data acquisition
electronics and the data analysis software.

In this paper we describe the instrument design philosophy, the
engineering design solutions and the manufacturing processes
adopted.  In a separate paper \cite{araujo06} we present full
performance Monte Carlo simulations for the final, as built,
instrument.

%%%%%%%%%%%%%%%%%%%%%%%%%%%%%%%%%%%%%%%%%%%%%%%%%%%%%%%%%%%%%%%%%%%%%%%%%%%%%%
\section{The ZEPLIN-III instrument}
\begin{figure}[ht]
\centerline{\epsfig{file=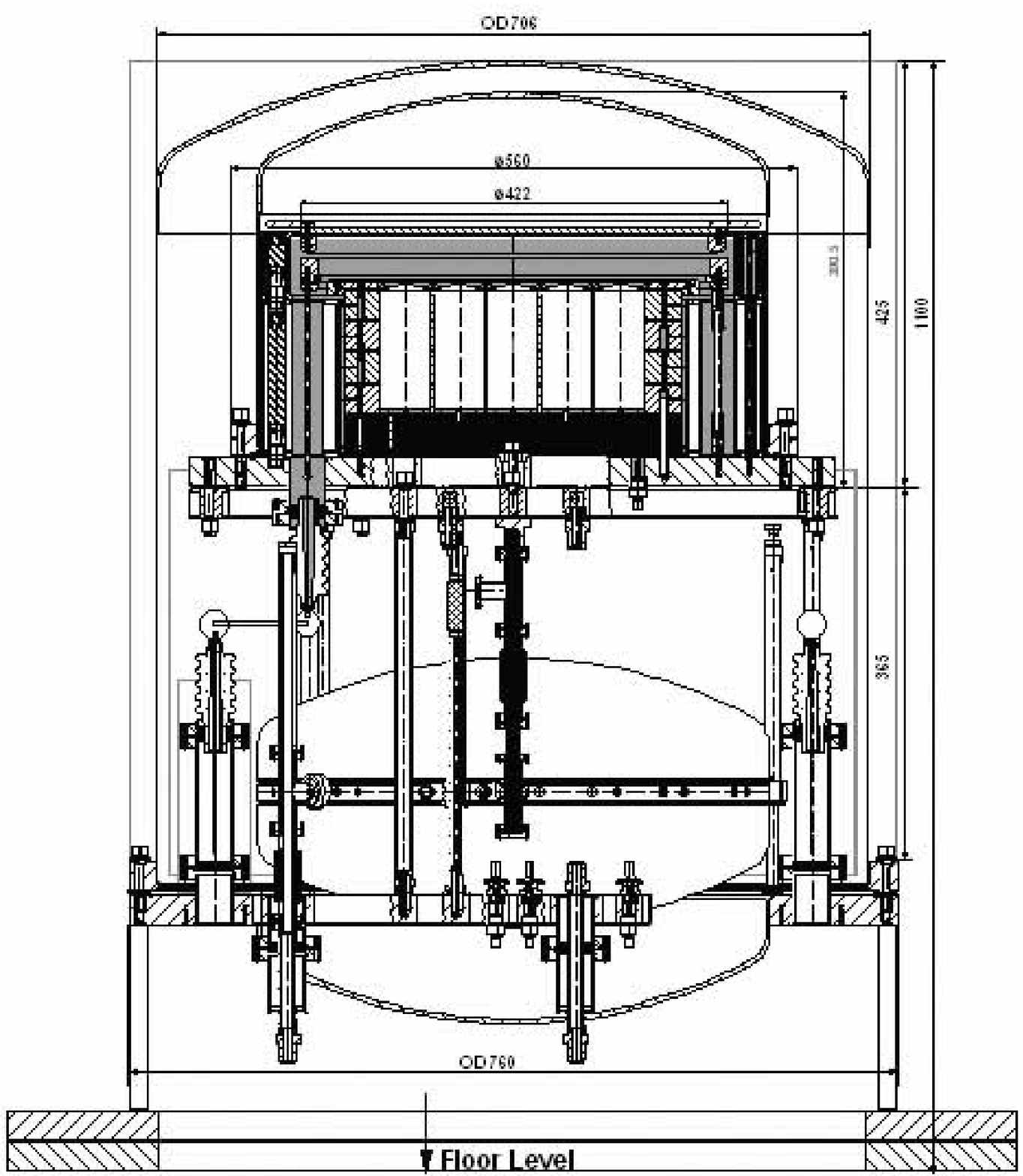,width=7.7cm}\epsfig{file=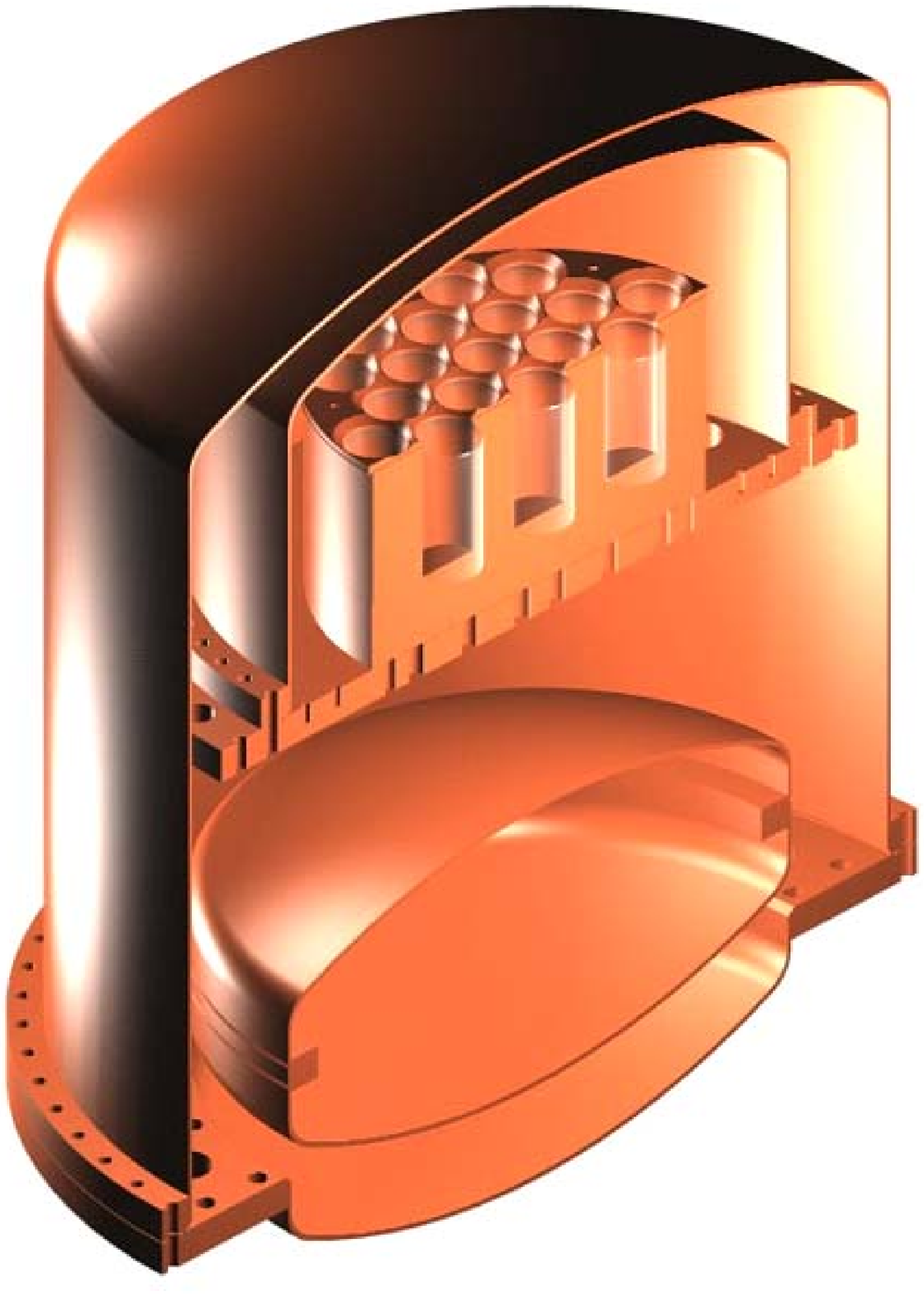,width=6.3cm}}
\caption{\small Cross-sectional views of the ZEPLIN III instrument
showing the key system design concepts.  The rendered CAD
representation shows the copper parts.} \label{z3}
\end{figure}

There are four important design requirements for a dark matter
detector: a low energy threshold, good particle discrimination,
3-D position reconstruction and a low background within the
fiducial volume. The ZEPLIN III approach, as shown in
figure~\ref{z3}, tries to push the boundaries of the two-phase
xenon technique to simultaneously achieve the best performance
possible in these four aspects.

ZEPLIN-III achieves a low threshold for the primary scintillation
by placing its photo-detectors, photomultipliers (PMTs) in the
liquid phase and by using a flat planar geometry.  Using PMTs in
the liquid removes two interfaces, both with large refractive
index mismatches and puts in an additional interface at which
total internal reflection also works to improve the light
collection for the primary scintillation.  The planar geometry
gives a large solid angle acceptance and lessens the dependance on
surface reflectivities. A low threshold for the
electroluminescence from the gas phase which provides the
secondary signal is achieved by using a high electric field in the
gas region to produce high levels of photon emission per electron
emitted from the surface and by using refraction at the liquid
surface to produce a `focusing' effect for the light onto the
immersed PMT array.

ZEPLIN-III achieves good particle discrimination between the
nuclear recoil signals expected from WIMPs and the electron
recoils from photon backgrounds by employing a two-phase design
which allows both scintillation and ionisation to be measured for
each event. The ratio of these two signals depends on the particle
species. The effectiveness of this discrimination depends on the
width and separation of the distributions for each species. It
turns out \cite{aprile06} that the discrimination is improved by
working at moderate electric fields which increases the separation
between the two distributions and improves the statistical
uncertainties of the ionisation signal.  Some discrimination
against nuclear recoil signals from neutron elastic scattering is
obtained by having good 3-D position reconstruction which can
identify the multiple scattering expected from the much higher
cross-sections for neutron scattering than for WIMP scattering.
Efficient measurement of the ionisation relies on achieving a long
lifetime against trapping for free electrons in the liquid.  This
requires ultra-pure xenon as free from electronegative impurities
as possible.  The target volumes must be constructed as high
vacuum vessels and a dedicated gas purification system is needed.

ZEPLIN-III achieves good 3-D position reconstruction by using an
array of 31 2" diameter photomultipliers.  These provide sub-cm
2-D spatial resolution in the horizontal $r$, $\theta$ plane.
Resolution in the $z$ co-ordinate at the $\sim50\,\mu$m level is
obtained from the timing between the primary and secondary
scintillation signals.

ZEPLIN-III achieves a low background partly by operation
underground and partly by using a very restricted range of
materials for its construction.  Although the PMTs are the largest
specific contributors to the background budget it is important
that careful attention is paid to all materials used as these
exceed the PMT mass by two orders of magnitude.  In addition it is
planned to eventually replace the PMTs by low-background versions
which are currently in development.

In the following sections we detail the design and manufacture of
the individual parts of the ZEPLIN III experiment.  These include
the target volume, the cooling system, the outer vacuum jacket,
the gas handling system, including the safety reservoirs, and the
data acquisition system.  In the final section we provide data
from surface commissioning tests which validate the key
performance parameters for ZEPLIN III

\section{The target volume}
\begin{figure}[ht]
  \centerline{\epsfig{file=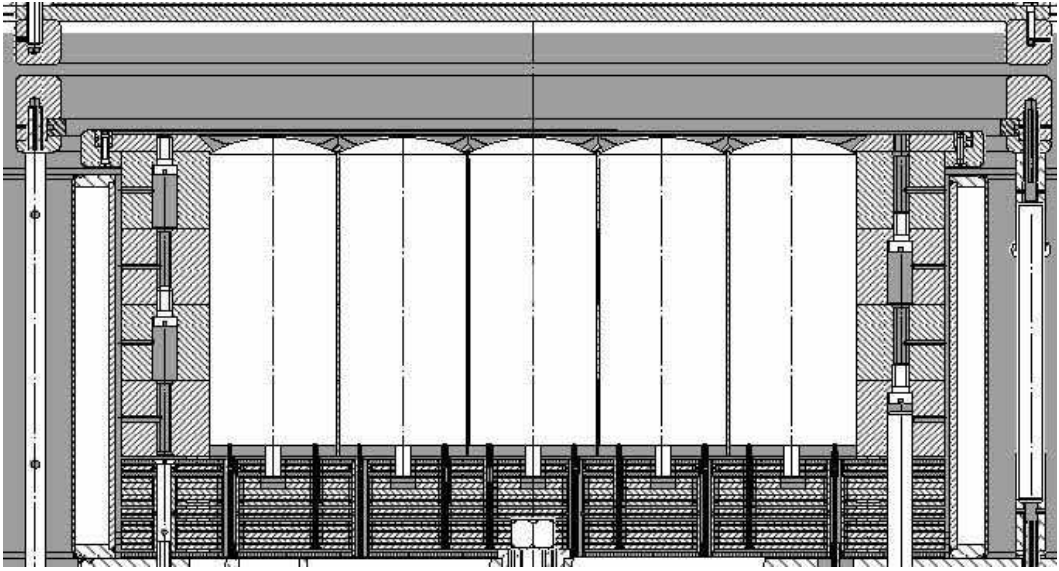,width=14cm,clip=}}
  \caption{\small Cross-sectional assembly drawing of the internal chamber volume of ZEPLIN III.}
  \label{target_volume}
\end{figure}

The detailed design of the inner components within the target
volume is shown in figure~\ref{target_volume}.

\subsection{The PMT array}
Inside the xenon vessel is the array of 31 PMTs, immersed in the
liquid phase, looking up to a $\simeq$40~mm-thick liquid xenon
layer on top of which is a 5~mm xenon gas gap. The figure shows a
cross-sectional view through a centre line passing through 5 of
the 52~mm diameter PMTs.  The others are arranged in a hexagonal
close-packed array with a pitch spacing of 54~mm.  A pure copper
`screen' has an array of 53~mm holes into which the PMTs fit. This
provides both light screening and electrical isolation between the
PMTs. It has an outer diameter of 340~mm and a height of 128~mm.
For ease of manufacture the total height of the `screen' was made
in four sections.  Each PMT hole through the copper `screen' has a
diameter of 53mm giving a 1~mm wall minimum thickness between each
PMT.  Two techniques were used to produce such thin wall section
through such a thickness of copper; wire erosion and boring. Both
worked but the boring produced a better surface finish.  Sitting
directly on top of the `screens' is another copper disc with holes
in it.  This time the thickness is 7~mm and the holes are finished
with highly polished conical sections to improve the light
collection; this plate is hence referred to as the `PMT mirror'.

\begin{figure}[ht]
  \centerline{\epsfig{file=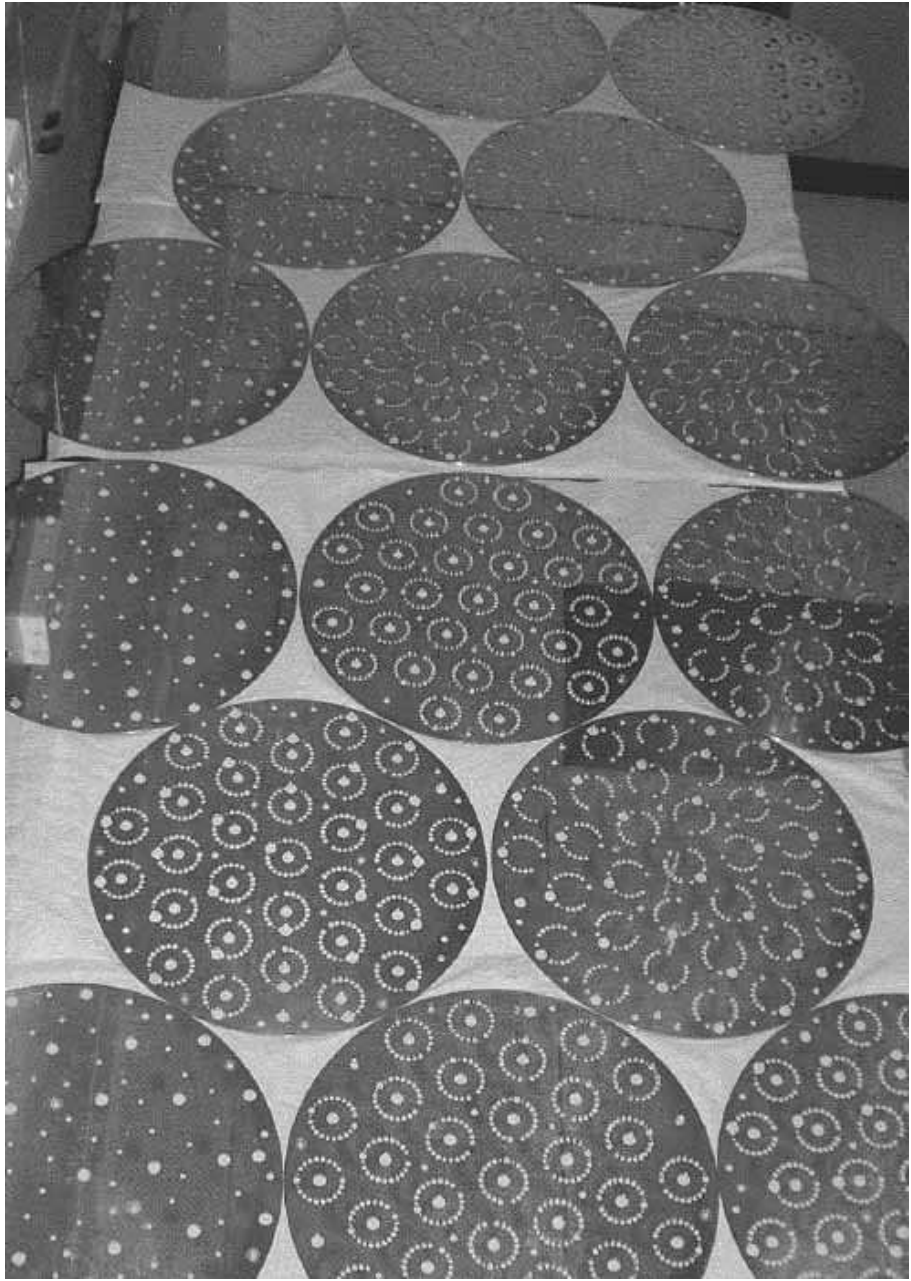,width=12cm}}
  \caption{\small The 16 2-mm copper plates used to make the internal PMT dynode interconnections.}
  \label{dynode_plates}
\end{figure}

\begin{figure}[ht]
\centerline{\epsfig{file=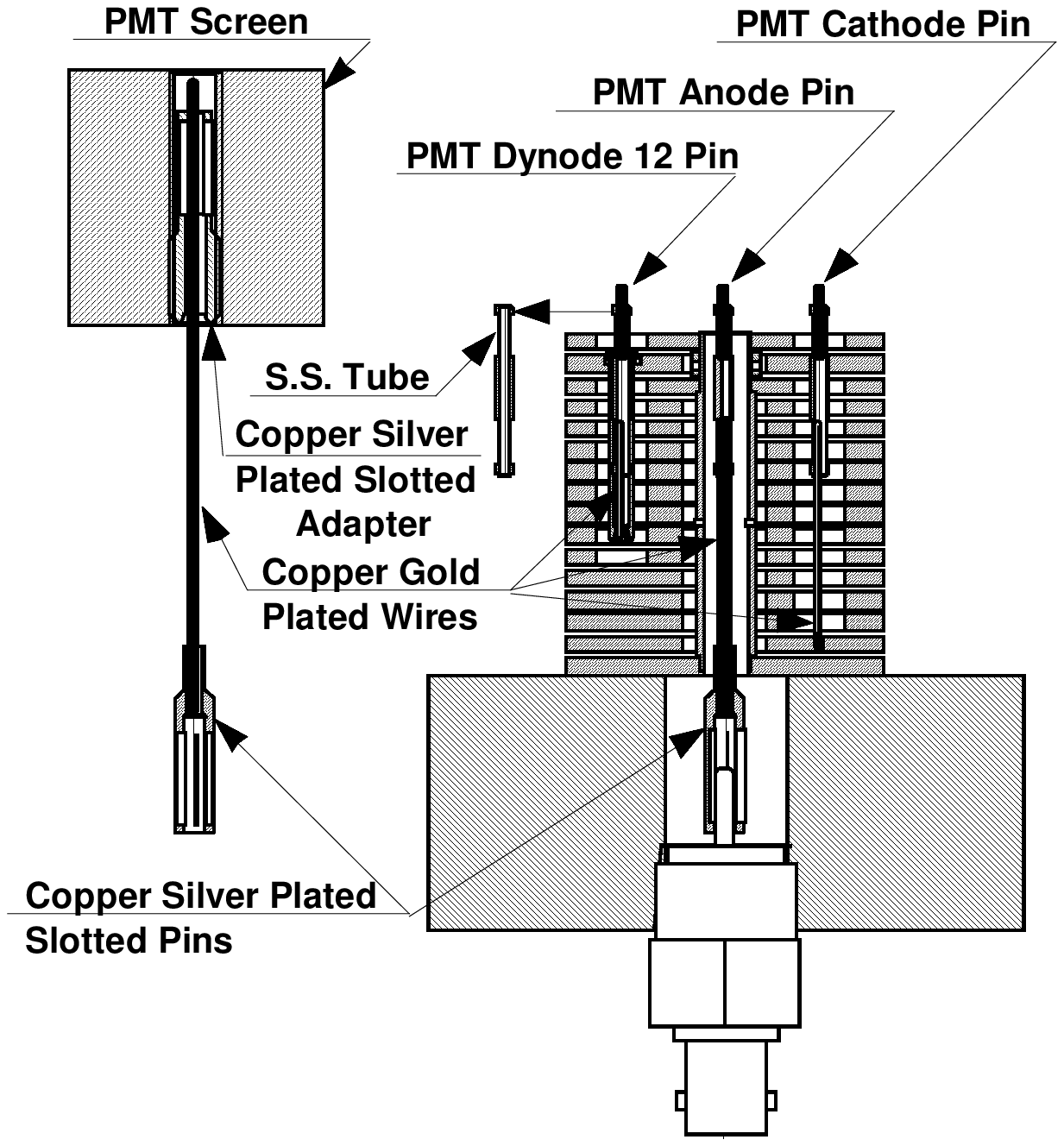,width=7.5cm}}
\caption{\small Various spring loaded contacts used to connect the
PMT pins to the copper plates.} \label{spring_contacts}
\end{figure}

Each PMT has 15 pins to which connections must be made (12
dynodes, anode, cathode and focus). However it would be
impractical to bring all 465 connections out through individual
UHV electrical feeds through the bottom thick copper flange.
Instead all the PMTs are run from a common high voltage supply and
dynode distribution system which reduces the amount of
feedthroughs to just 47. The corresponding dynode pins on each PMT
are connected together using a stack of 16 thin copper plates,
held apart with small quartz spacers, below the PMT array. Each
2mm thick plate has a different pattern of holes (see
figure~\ref{dynode_plates}) allowing connection to each pin in
turn whilst the others pass through with clearance. Connections
between the copper plates and the PMT contacts were done by first
cold welding a pin into the copper plate and then using spring
loaded tubes to join the two pins together (see
figure~\ref{spring_contacts}).   The pins used in the copper
plates were made in copper with a gold coating and these were
inserted into tight fitting holes in the plates using a drill
press.  The spring contacts were made from stainless steel tubing
with reduced wall sections and slots.  These contacts provide
enough friction for retention of the PMT against buoyancy forces
during immersion in liquid xenon. Connection between each plate
and its single UHV coaxial feedthrough was again made by a direct
spring loaded tube but with the addition of gold-plated copper
wires with silver-plated copper adaptors to provide the extensions
between end contacts. The anode connection from each PMT is
brought out separately on a dedicated coaxial UHV feedthrough in a
similar way.  The specific arrangement of the 16 copper plates can
be seen in figure~\ref{spring_contacts}.  The upper and lower
plates are connected to ground.  The second lowest plate is
connected to the PMT cathodes, the next 11 are connected to
dynodes 1 to 11 in turn. Above that there is then another grounded
plate and between this and the top ground plate is dynode 12.  The
two grounded plates either side of the dynode 12 plate
deliberately provide both extra capacitance to ground for that
dynode and prevent cross-coupling with other connections. Copper
tubes provide shielding along the run of each anode output
connection.  Shielded cables pass across the outer vacuum jacket
space to connectors in its base plate.  A single external voltage
divider chain is used to provide all the common dynode voltages.
To ensure reasonably well matched gains when running from a common
HV supply, PMTs were procured with gains within prescribed limits.
Once selected the batch of 35 PMTs (ETL D730/9829Q) was tested and
calibrated at low temperature with Xe scintillation UV light prior
to installation in the detector \cite{araujo04}. The PMTs were
customised specifically for ZEPLIN III in two ways: firstly a
conductive pattern of so-called `fingers' was deposited on the
inside of the window to avoid saturation at high count rates, and
secondly to provide a modified pin-out arrangement to facilitate
the use of the copper interconnection plates.  The PMTs are
operated with the cathode at ground potential.

\subsection{The electric field}
Proper operation in two-phase mode requires that there be a
sufficiently high electric field in three distinct regions. In the
active volume of the detector the electric field helps to separate
ionisation charge released from the track of the interacting
particle before it can recombine. This field must be directed such
that the electrons start `drifting' towards the liquid surface.
Hence the field in this first region is called the `drift' field.
The second critical region is at the liquid/gas interface.  Here
the field in the liquid must be high enough to efficiently extract
the electrons into the gas phase. This not only increases the
signal strength but also prevents charge build-up at the surface.
This field is called the `extraction' field.  Finally in the gas
phase the field must be high enough for the accelerated electrons
to produce excitation in the gas atoms. The excited atoms then
form excited dimers followed by dissociative radiative emission in
the usual way, which produces the signal seen by the PMTs.  This
last field is called the `electroluminescence' field.  These three
fields can either be produced by setting up a segmented electrode
structure producing distinct regions, as is done in ZEPLIN
II~\cite{smith06}, or, as in the case of ZEPLIN III, a single pair
of outer electrodes can be used to produce all three at once.  The
advantage of the latter is the absence of any physical electrode
structure in the liquid which could then be a source of background
and/or feedback. However it does mean that a single much higher
individual voltage is required and the fields can not be
controlled independently. The two electrodes used are the solid
flat plate (`anode mirror') above the gas gap and a wire plane
(`cathode grid') 40~mm below it in the liquid. The 8-mm top plate
is made from copper and its bottom surface has been lapped using
optical techniques and left highly polished. Up to 40~kV can be
applied between the two `electrodes'.

\begin{figure}[ht]
  \centerline{\epsfig{file=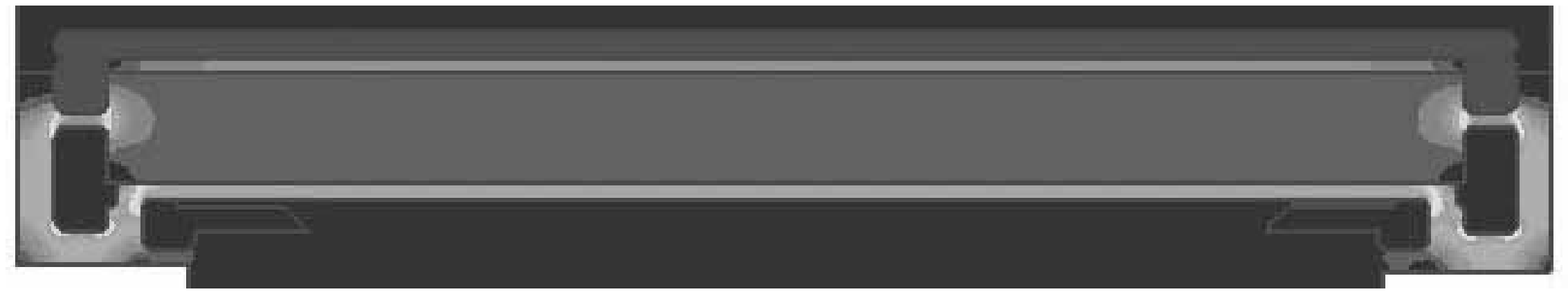,width=8cm,clip=}\epsfig{file=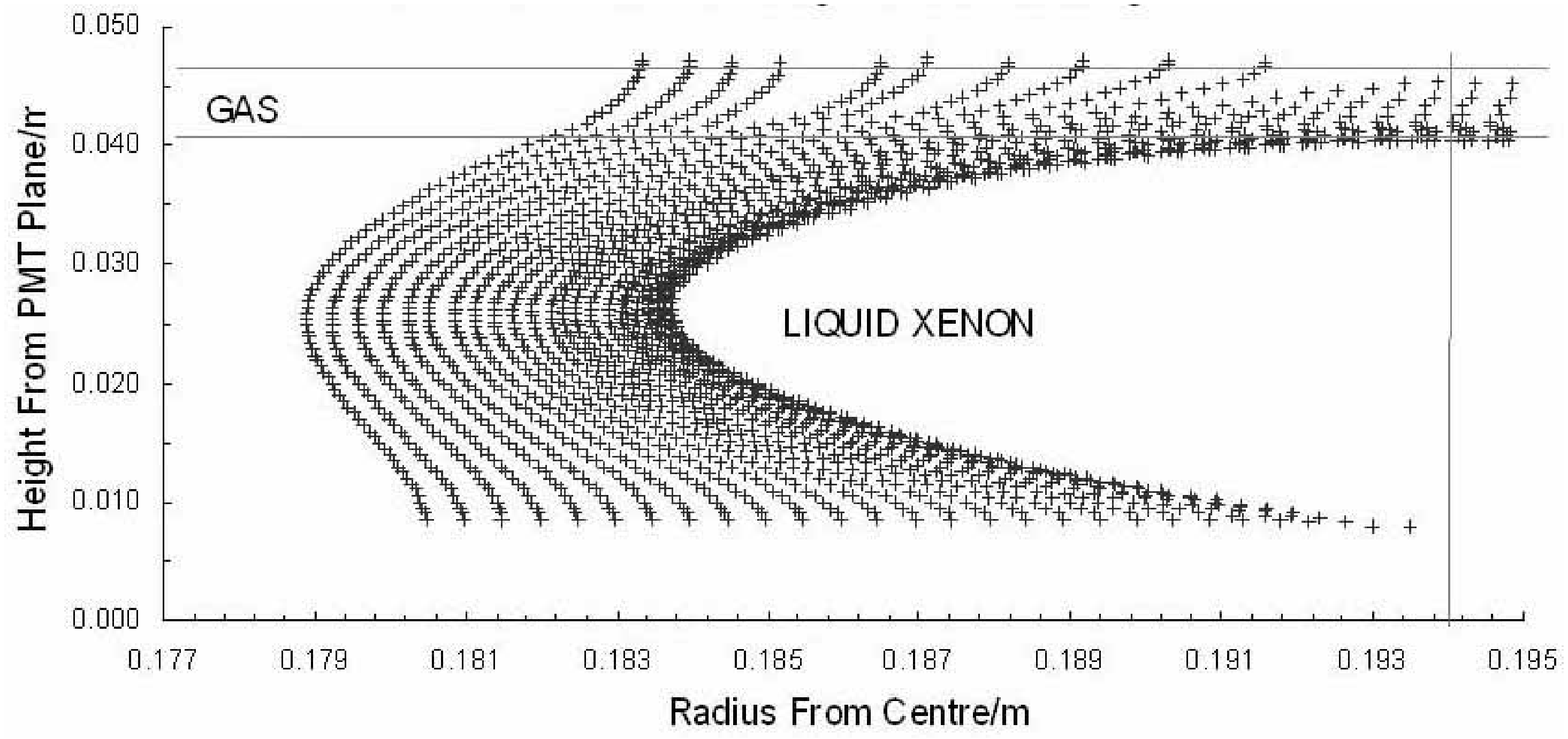,width=6cm,clip=}}
  \caption{\small Electric field distribution within the target volume as computed using ANSYS~\cite{ansys}.
  On the right is shown an expanded view of the drift paths of electrons near the right-hand gap between the two electrodes.}
  \label{efields}
\end{figure}

A second wire grid (`PMT grid') is located 5~mm below the cathode
grid and just above the PMT array.  This defines a reverse field
region just above the PMTs which suppresses secondary signals from
low-energy background photons from the PMTs and also helps isolate
the internal PMT photocathode fields from the external high
electric field.  The diameter of the electrode structure is
$\sim40\,$~cm, whilst that of the PMT array is 34~cm. The fiducial
volume will be defined by a combination of primary to secondary
timing and position recovery from the PMT hit pattern, and it will
be well inside the PMT array diameter.  This ensures that the
electric field will be very uniform over the fiducial volume
region. Field and electron trajectory simulations, produced using
ANSYS \cite{ansys}, are shown in figure~\ref{efields}.

\begin{figure}[ht]
  \centerline{\epsfig{file=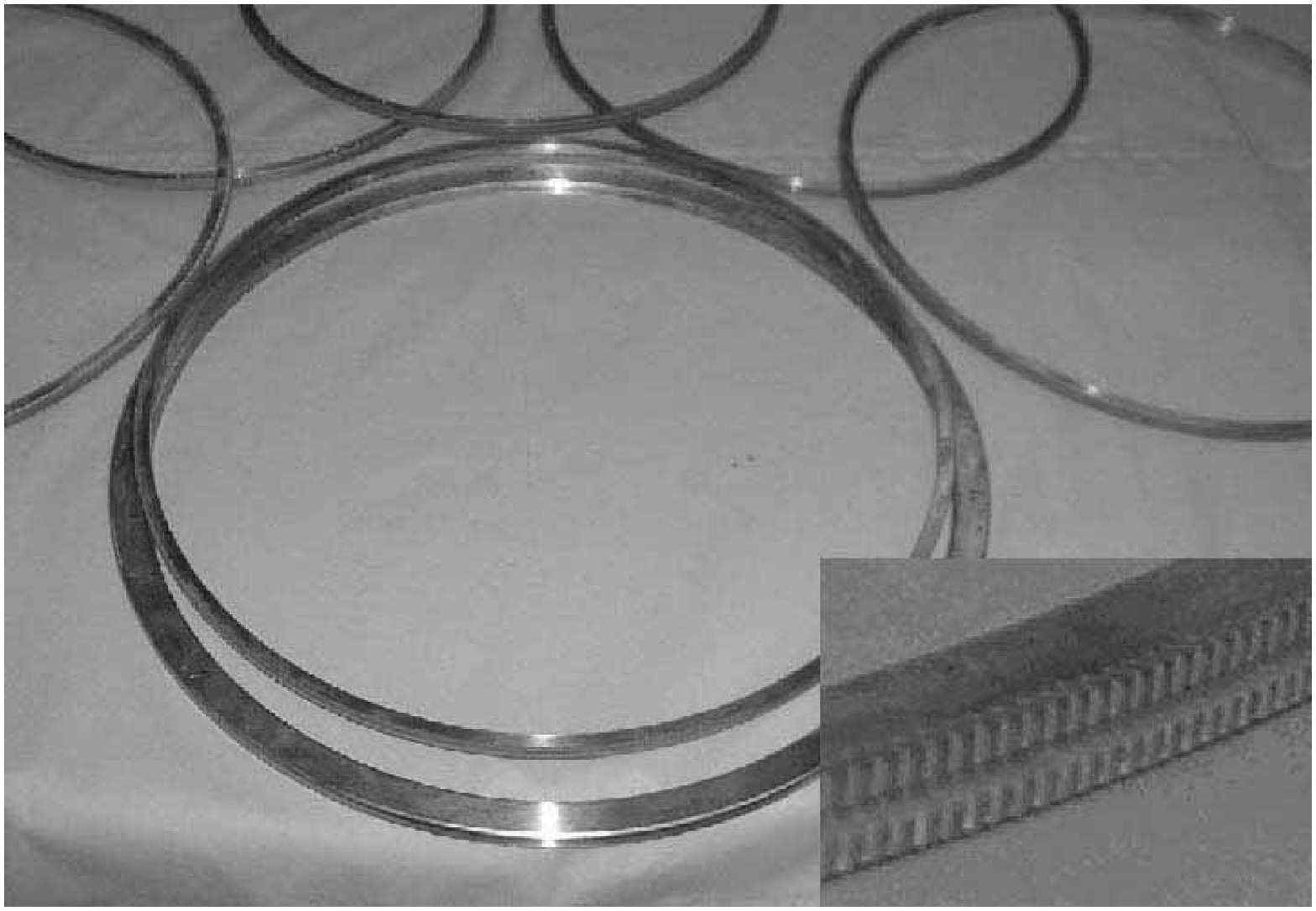,width=7cm}
  \epsfig{file=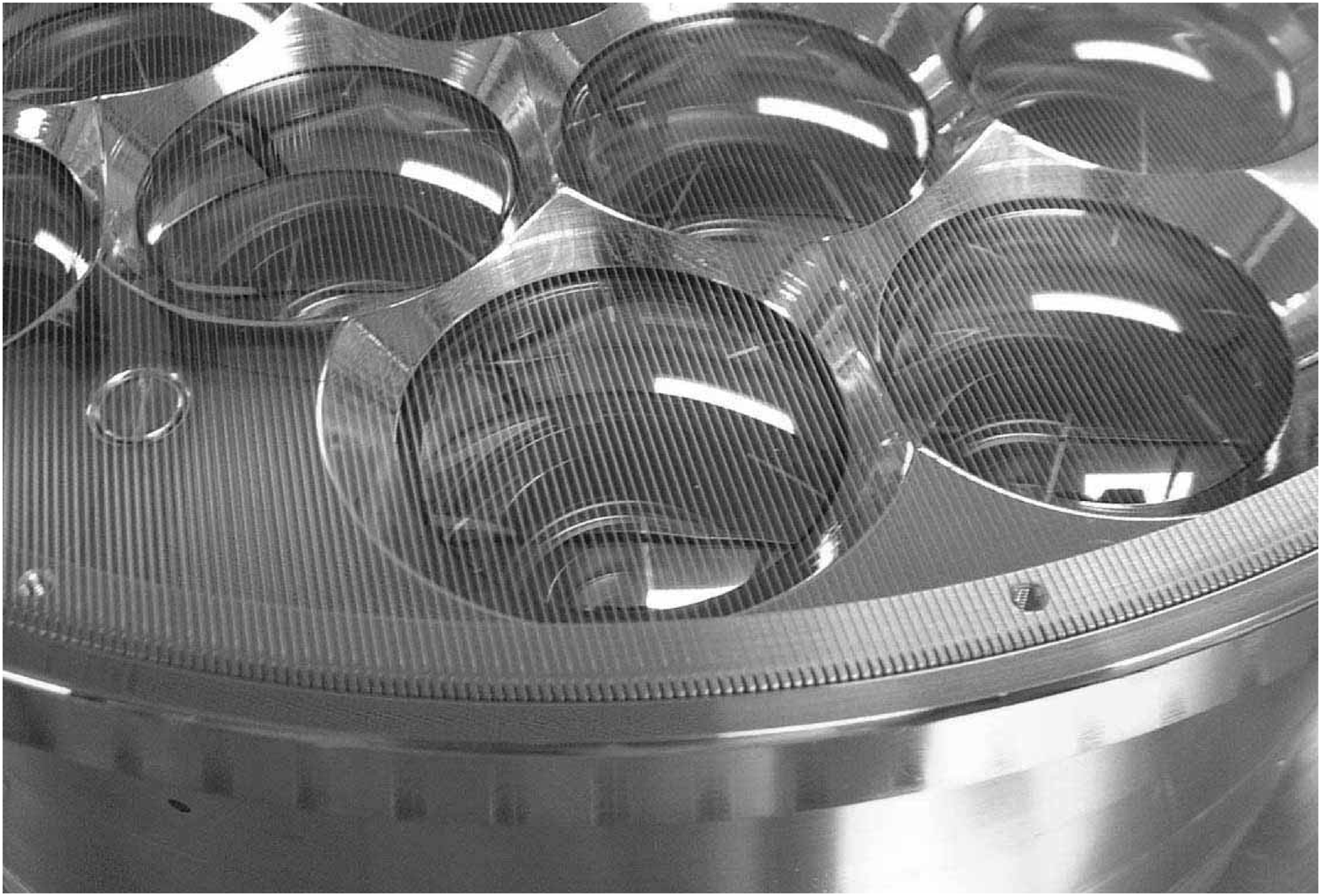,width=7cm}}
  \caption{\small On the left are the copper formers for the wire grids.
  The inset detail shows an expanded view in which the slots cut to control to wire positioning can be seen.
  On the right is a view of the assembled PMT array in which the PMT grid can be seen.}
  \label{grids_plus}
\end{figure}

The stainless-steel wire grids were strung from continuous lengths
of $100\,\mu$m diameter wire wound around copper formers. The
position of each wire was controlled by slots machined into the
formers (see figure~\ref{grids_plus}).  The wires were tensioned
using two techniques. Firstly the formers were elastically
deformed whilst the wire was wound and secondly the winding jig
tensioned the wire as it was wound. Once the grid winding was
complete the wire was anchored and the formers were then released
from their restraining jigs.

Some consideration was given to whether the anode mirror should be
coated to enhance its reflectivity.  The performance of polished
copper is quite uncertain at VUV wavelengths, depending on the
surface finish, oxidation state and possible LXe condensation onto
the cold surface in the gas phase. Only a single measurement has
been found, indicating $R$=25\% for normal incidence for a
clean-cut surface \cite{handbook}.  However the simplicity of
leaving this surface as is, the uncertainty of using coatings in a
high-field application  and the desire not to compromise the
spatial reconstruction argued for not using any coating.

\subsection{The xenon transport system}
\begin{figure}[ht]
\centerline{\epsfig{file=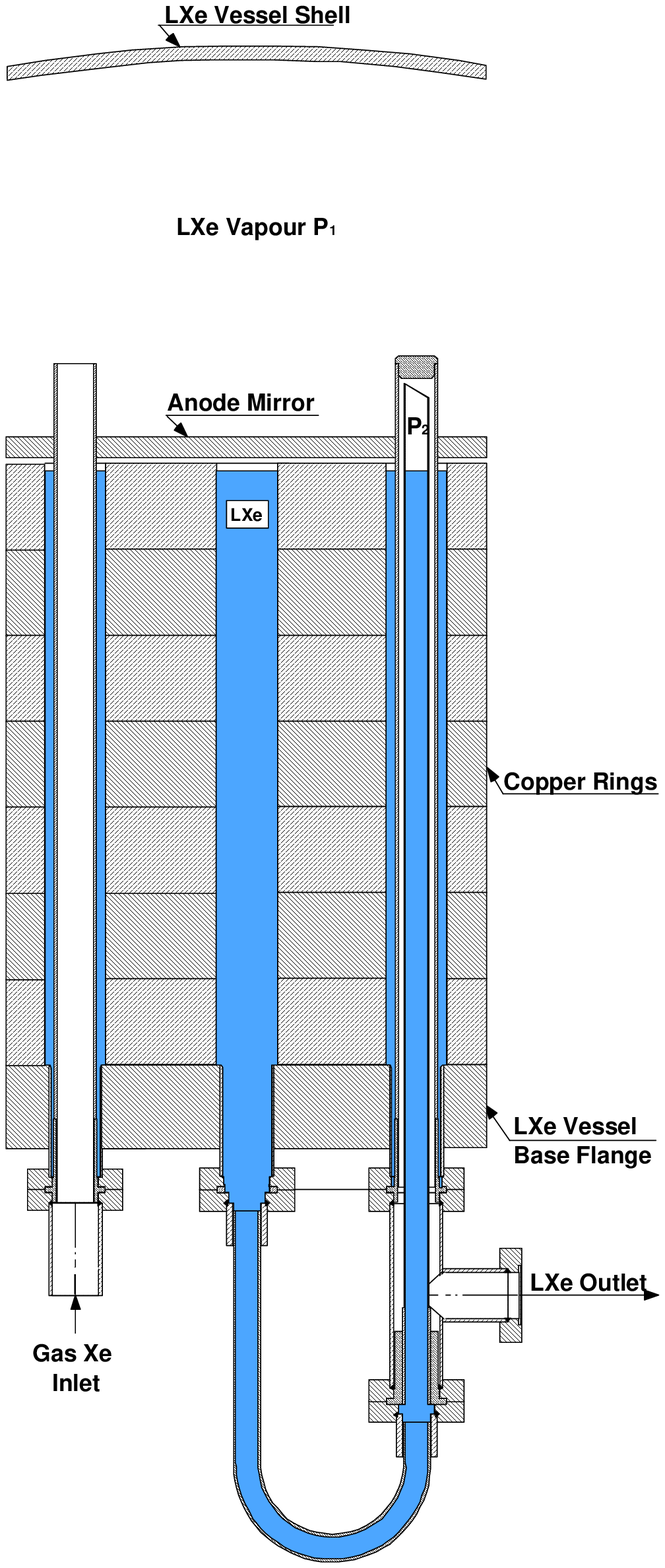,width=6cm}}
\caption{\small Arrangement of the pipework used for xenon
transfer in and out of the target vessel.} \label{pipes}
\end{figure}

Two copper access pipes are included for movement of xenon in and
out of the target vessel (see figure~\ref{pipes}). One surfaces
above the liquid level in an unconfined volume as is used as a
`Gas Inlet'. The second has a double tube structure with an open
ended inner pipe connected directly to the main xenon liquid
volume, and an outer pipe which vents to the outside through the
`LXe Outlet'. The outer pipe is sealed at the top and the inner
opens above the liquid surface and essentially allows a `syphon'
action during emptying. Transfer of xenon in and out of the target
vessel is independent of the cooling system.

\subsection{The target vessel}
The containment vessel for the xenon target must perform as both a
high vacuum vessel, for purity reasons, and a pressure vessel for
safety reasons.  The pressure vessel design was done following the
relevant British Standard (BS5500:1997). This safety standard
dictates the cylindrical wall, dome and bottom flange thicknesses
which are dependent on the material and processes used.  The
vessel was required to be certified to 6\,bar absolute.  The
material of choice was determined by requiring the product of
total mass times radioactive content be a minimum. Added to this
prime requirement was then the need for the material to be
suitable for manufacture of the vessel.  OFHC copper type C103 was
selected.  This required 4\,mm wall thickness on the cylindrical
sections, 3\,mm for the spun domes and 25\,mm for the flat bottom
flange. To minimise the likelihood of inclusion of any impurities
electron-beam welding was used throughout and the number of welds
was kept to a minimum. In particular the cylindrical section was
rolled in one piece. Stainless steel parts were used for some
specialist components which would have been very difficult to make
out of copper, such as vacuum knife edge flanges and vacuum HV
feedthroughs for which commercial parts were used. Where necessary
these stainless parts were also electron-beam welded to the
copper.  Welding techniques adapted to our requirements were
developed by The Welding Institute, UK\cite{twi}, in close
cooperation with us.  This included setting the welding parameters
and optimising the structural/thermal design of the weld joints.
All safety critical welding was done by certified processes and
copper witness plates were used to ensure proper and complete
breakthrough as all welds were required to show full-depth
penetration.  Special jigging was required to hold all seams for
welding securely in place during the process.
%Figure~\ref{welds} shows some examples of joint designs and weld finishes.
On completion all joints were
leak-tested down to the level of $\sim10^{-10}\,$mbar.l.s$^{-1}$.
%\begin{figure}[ht]
%\centerline{\epsfig{file=missing_figure.eps,width=12cm}}
%\caption{\small Weld joint designs and weld finishes from the
%inner vacuum vessel.} \label{welds}
%\end{figure}

\begin{figure}[ht]
\centerline{\epsfig{file=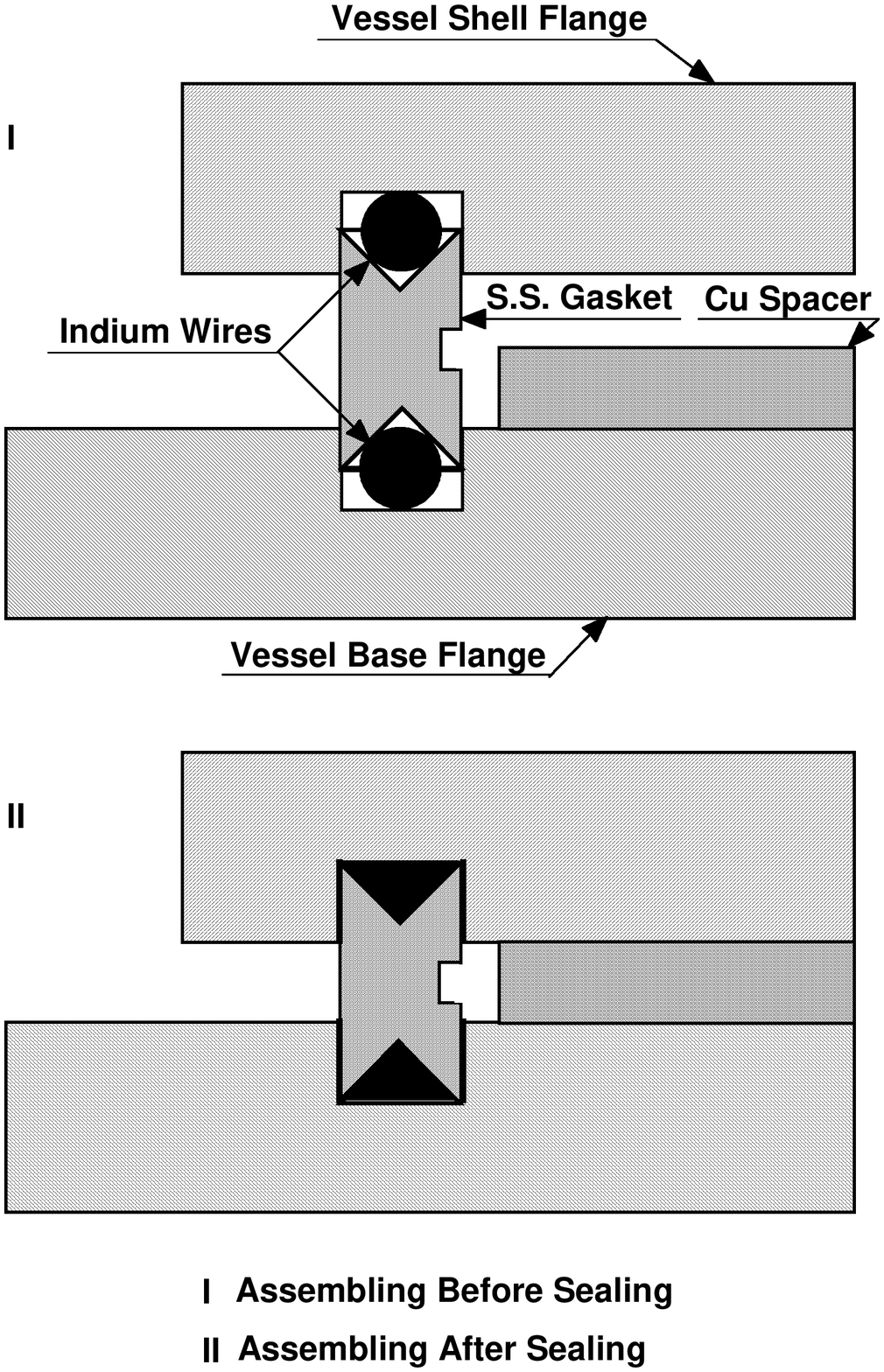,width=7cm,clip=}}
\caption{\small Vacuum seal between the cylindrical wall section
and the bottom flange before (I) and after (II) sealing.}
\label{flange_gasket}
\end{figure}

The electrical feedthroughs for the PMT dynode connections were
fitted in with screw threads with thick indium coated onto them
using an ultrasonic soldering iron.  The demountable vacuum seal
between the cylindrical section and the bottom flange was done
using a stainless steel gasket with double knife edges and an
indium wire at both copper surfaces (see
figure~\ref{flange_gasket}).
All copper parts were cleaned, starting with a coarse hand
polishing with stainless wire wool, a fine polishing with copper
wire wool, an ultrasonic bath using 2\% CITRANOX ~\cite{alco} in
de-ionised water and a high-pressure wash using pure water.  The
polishing phase was done using a powered rotation table specially
built for the purpose and polishing was always applied along the
line of existing machining marks.

\section{The cooling system}
\begin{figure}[ht]
\centerline{\epsfig{file=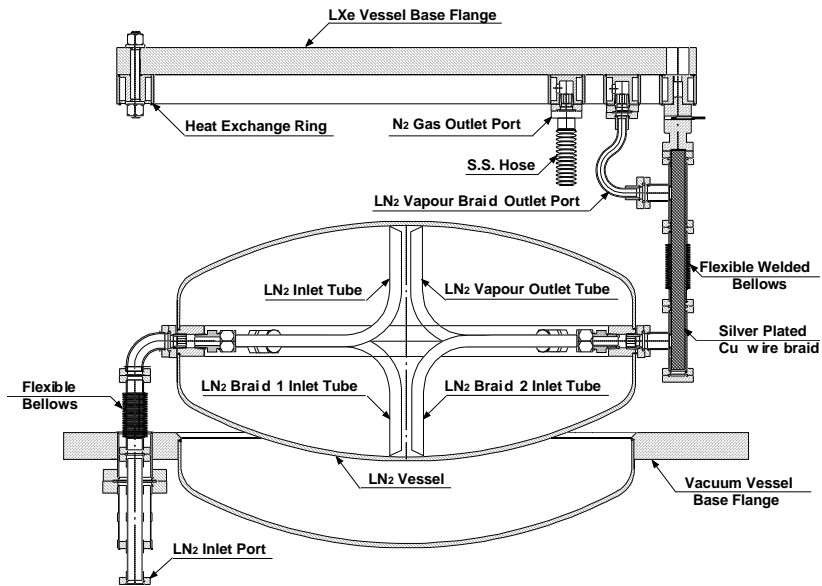,width=8cm,angle=-90,clip=}}
\caption{\small Thermal control system elements.} \label{thermal}
\end{figure}

Cooling is done using liquid nitrogen (LN2).  The internal
reservoir, located under the target vessel, holds 36\,litres.
There are two thermal links between this reservoir and the target
vessel (see figure~\ref{thermal}). The first link is a conduction
path provided by flexible bundles of thick copper wires thermally
anchored to a hollow copper cooling flange (`heat exchange ring')
attached to the underside of the target vessel. The flexibility
helps decouple acoustic/mechanical noise in the LN2 reservoir from
the LXe chamber. The other end of the bundle dips into the liquid
nitrogen and the thermal impedance depends on the depth of the
liquid.  The bundle is welded and polished at both ends for good
thermal matching.  A second thermal path is provided by a direct
connection between the nitrogen reservoir and the hollow cooling
flange. This allows liquid or boil-off gas to be used as
additional coolant and provides the means for active thermal
control with minimum cryogen usage, which is important during
stand-alone operation underground.

\begin{figure}[ht]
\centerline{\epsfig{file=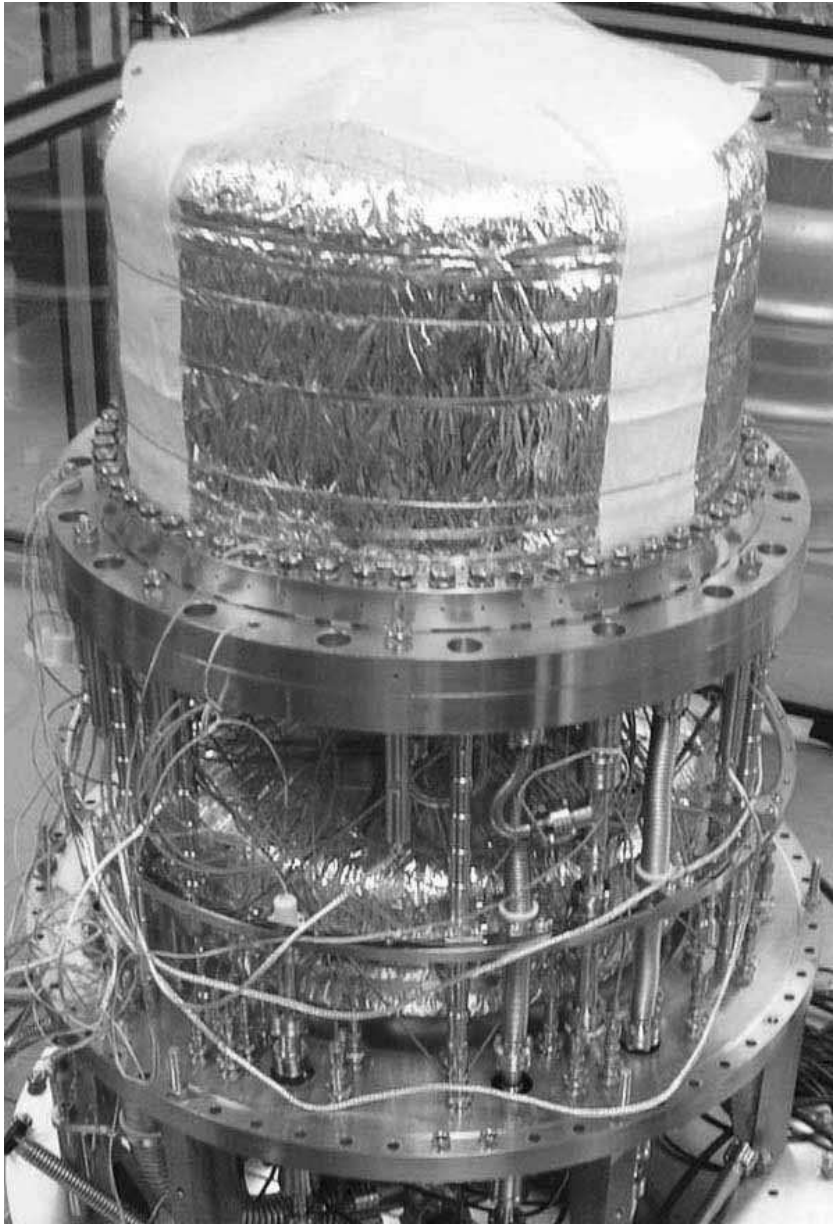,width=11cm}}
\caption{\small The assembled instrument without its vacuum jacket
giving a view of the target vessel (top) and liquid nitrogen
reservoir both covered with thermal insulation.}
\label{vessel_nitrogen}
\end{figure}

Three external pipes are connected to the cooling system. The
first is the liquid nitrogen delivery line and this terminates
inside the reservoir close to the top.  A second pipe also opens
to the top of the reservoir, whilst a third pipe connects to the
delivery to the cooling flange. The last two pipes are fitted with
control valves which regulate the internal pressure and the flow
rate through the hollow cooling flange. During initial cool-down
the flow through the cooling flange is increased to allow bulk
liquid flow into it. Once cold, the gas flow through the cooling
flange provides a fine temperature control mechanism whilst the
copper cable bundles provide the main thermal link balancing the
average heat load. The heat load is reduced by the use of thermal
insulation around both the target vessel and the nitrogen chamber
(see figure~\ref{vessel_nitrogen}).  The nominal operating
temperature is around $-100{\rm^oC}$ and the heat load is, as
expected, $\sim40\,$W, giving a design hold-time between refills
of $\sim$2\,days.

\section{The outer vacuum jacket}
\begin{figure}[ht]
\centerline{\epsfig{file=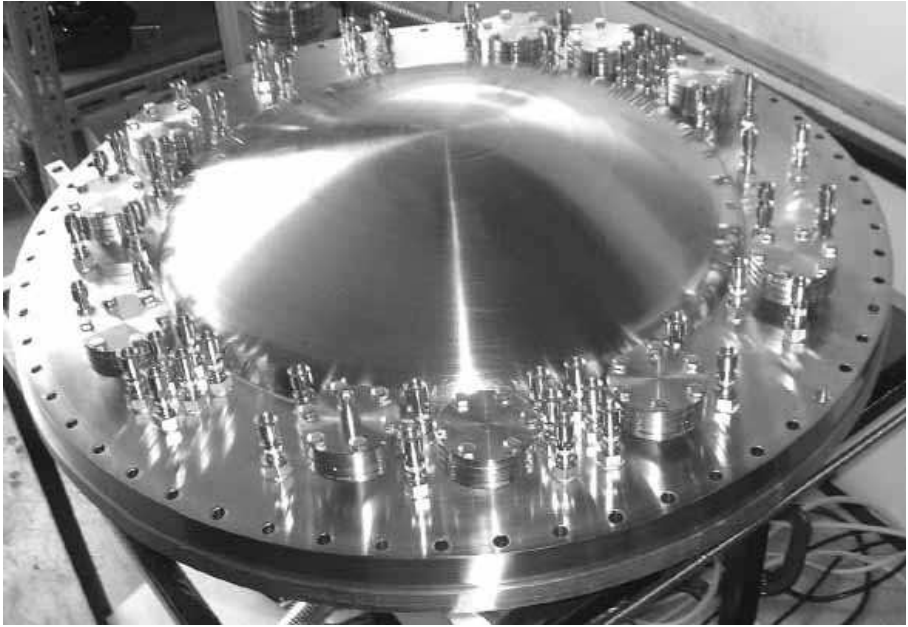,width=12cm}}
\caption{\small The bottom flange of the vacuum jacket showing the
inner domed section and the arrangement of feedthroughs and ports
around the outer skirt.} \label{bottom_flange}
\end{figure}

The design principles for the vacuum jacket were much the same as
for the target vessel, except that the pressure rating was reduced
to 4.3 bar absolute. The safety standard for pressure vessels
dictated the material thicknesses and process standards and the
same attention to background and cleanliness was imposed. Hence
OFHC copper was used, with electron beam welding and minimisation
of the number of seams; the cylindrical section of this larger
vessel was also made from just one rolled plate. The bottom flange
has an included domed section and the vacuum seals were all done
in the same way as for the target vessel.
Figure~\ref{bottom_flange} shows the underside view of the bottom
flange.

\section{The gas purification system and safety reservoirs}
\begin{figure}[ht]
\centerline{\epsfig{file=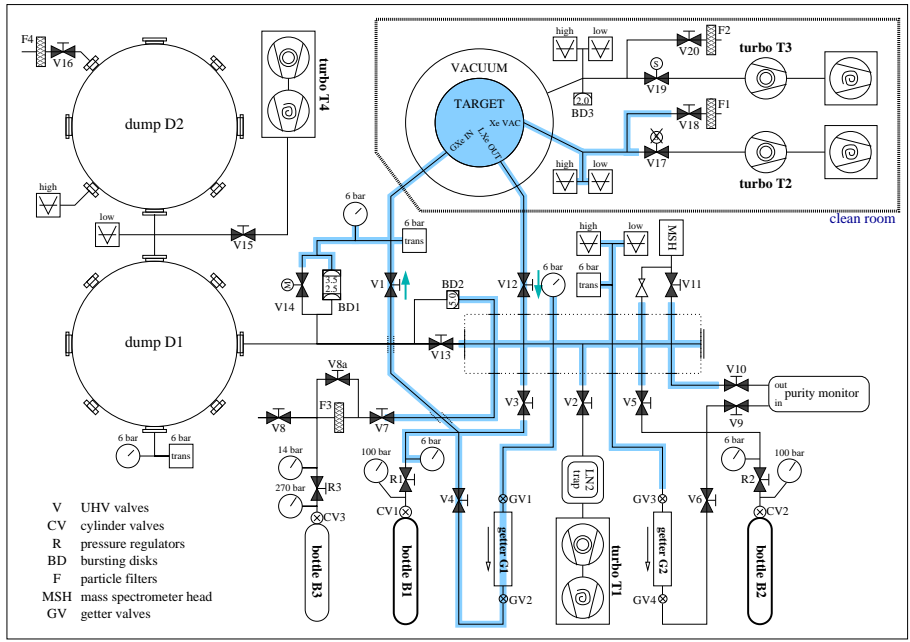,width=12cm}}
\caption{\small Schematic diagram of the gas purification system.}
\label{gas_purification}
\end{figure}

Figure~\ref{gas_purification} shows a schematic of the xenon gas
purification system.  The main requirement is to be able to remove
electronegative contaminants which will prevent the ionisation
electron drift and suppress the secondary signal. Typically this
requires liquid xenon purity down to the parts per billion level,
beyond that available through commercial purchase. In addition the
level of radioactive krypton needs to be kept as low as possible
as the beta-decay of $^{85}$Kr gives a continuum energy deposit
down into the level expected from elastic scattering of WIMPs. An
all-metal bakeable gas system has been used.  The system is pumped
by a combination of oil-free scroll and turbo-molecular pumps. The
xenon gas is contained in two large stainless steel cylinders
fitted with high purity all-metal UHV valves and regulators. These
two cylinders stand in cooling jackets allowing them to be cooled
to liquid nitrogen temperatures.  Two SAES\cite{saes} getters are
used. Fine particle filters ($0.5\,\mu$m) are fitted to all gas
delivery lines. The gas system is fitted with a mass spectrometer
which is used both for helium leak testing and residual gas
analysis. The base vacuum attainable in the system is $\sim
10^{-8}$~torr, dominated by H$_2$; a partial H$_2$O pressure of
$\sim 10^{-10}$~torr was achieved prior to the xenon input. The
detector itself is connected without valves to a port on the main
volume of the gas purification system. Another port is connected
to the large volume safety reservoirs with only a burst disk
between them. This is not only to guard against the safety risk
associated with catastrophic failure of the target vessel under
overpressure, but also to avoid loss of xenon.  The two gas
cylinders contain 50\,kg of xenon supplied by ITEP from stock
collected from underground sources between 20 and 40\,years ago.
This xenon is expected to have a very low radioactive krypton
content. A final component of our gas purification system is a
novel portable chamber for electron lifetime measurements which
will be described elsewhere\cite{walker06}.

\section{The data acquisition system}
The 31 PMT signals are fed into wideband amplifiers and split into
a dual dynamic range data acquisition system (DAQ). This ensures
sensitivity to very small primary scintillation signals containing
only a few photoelectrons (phe) as well as to large secondaries
without saturation. All 62 channels are sampled at 500~MS/s by
8-bit AQIRIS digitisers.  For the collection of `dark matter' data
a PMT gain of $2\times 10^5$ will be used. Such a low gain should
avoid internal PMT saturation effects following very large
secondary scintillation signals. Wideband amplifiers add
electronic gain in two stages. The first stage is (x10) with a
noise referred to the input of 30~$\mu$V\,rms.  They then feed
into adjustable attenuators which are used to equalise the single
photoelectron response for each PMT. The outputs from this stage
then feed into the 31 low-gain digitisers as well as into the next
stage x10 wideband amplifiers. The high and low-gain input
channels thus have a factor of 10 gain difference which can be
further expanded by adjusting the full-scale ranges on the
digitisers. A simple threshold trigger signal is derived from a
summing amplifier, with inputs from all PMTs, fed into a
discriminator whose output provides an external trigger for the
AQIRIS digitisers.  This trigger can not differentiate between
primary and secondary scintillation signals. A more sophisticated
trigger using a time to amplitude converter can provide a width
measure and differentiate the very short primary scintillation
signals ($\sim30\,$ns time constant) from the much more extended
secondary scintillation signals ($\sim1\,\mu$s duration). The
maximum delay between primary and secondary scintillation signals
in ZEPLIN III is $\sim17\,\mu$s. A LINUX-based software
application reads out the digitiser crates. A FIFO-type memory
buffer, accessed independently by two CPUs for data transfer and
write-out, reduces the overall dead time. An acquisition rate of
100 events/s can be sustained.

\section{Commissioning cool-down tests}
The first cool-down test was designed to verify the thermal
control system and to test out the PMT array.  For this test the
anode and cathode electrodes were replaced by a copper plate
located just 8~mm above the PMT array. 31 $^{241}$Am radioactive
sources were vacuum-sealed into this plate with a thin copper foil
overlay to prevent leakage of radioactivity and to stop
$\alpha$-particles from interacting in the xenon.  These then
provided a source of low-energy (mainly 59.6~keV) photons.  For
subsequent cool-down tests the radioactive sources had been
removed and the full electric field system installed in its final
configuration.
\subsection{Cooling system}
\begin{figure}[ht]
\centerline{\epsfig{file=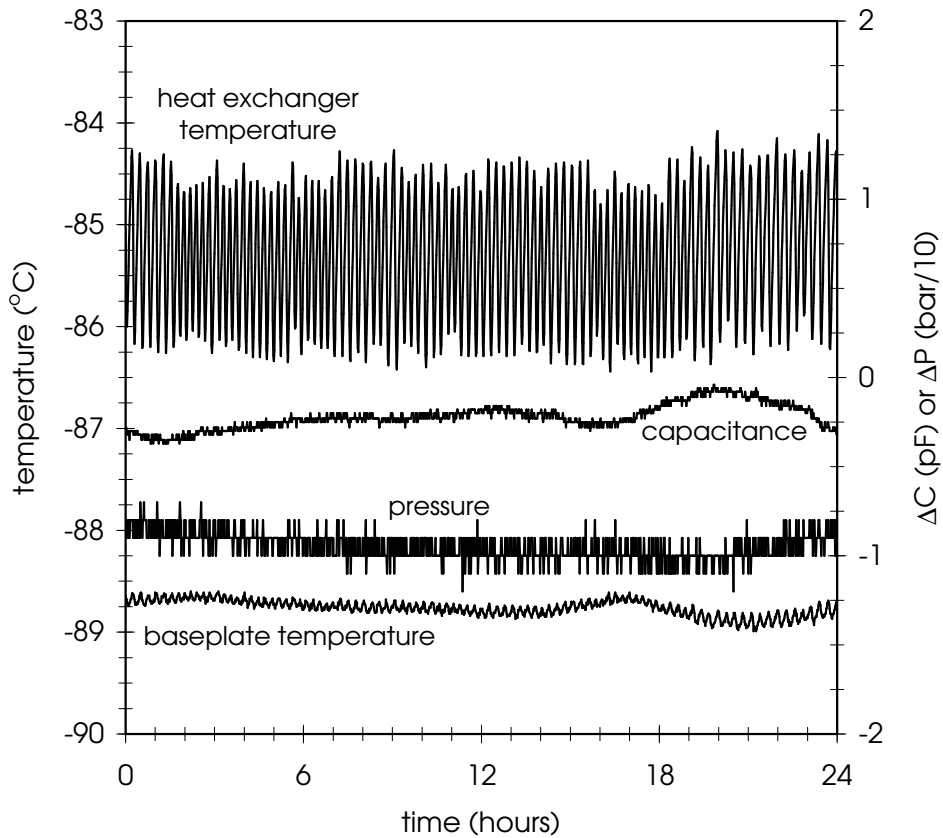,width=13cm}}
\caption{\small Some key system engineering parameters monitored
over a 24\,hour period.} \label{stability}
\end{figure}

The cooling system performance during the first cool-down was as
expected.  The initial cool-down period used 200\,litres of liquid
nitrogen and progressed at $\sim\,5~^{\rm o}$C/hour.  An array of
temperature sensors was used to monitor critical points within the
instrument.  One of these, on the lower face of the cooling flange
on the bottom of the target, is used as the control temperature
and its reading is compared with a set temperature in the
controller to automatically operate two valves: one which exhausts
straight from the gas volume of the nitrogen reservoir, and one
which exhausts through the cooling ring.  Once down at the nominal
operating temperature ($\sim -100~^{\rm o}$C) the temperature of
the target vessel is stable to better than $0.2^{\rm{o}}$\,C and
the liquid nitrogen usage drops to $\sim20\,$litres/day as
expected. Figure~\ref{stability} shows some key engineering
parameters monitored over a 24 hour period during the second
cool-down test. The upper trace is from the temperature sensor on
the cooling flange and the periodic behaviour is due to the
control system. The lower trace is then the temperature of the
base plate of the target vessel itself.

\subsection{PMT array}
\begin{figure}[ht]
\centerline{\epsfig{file=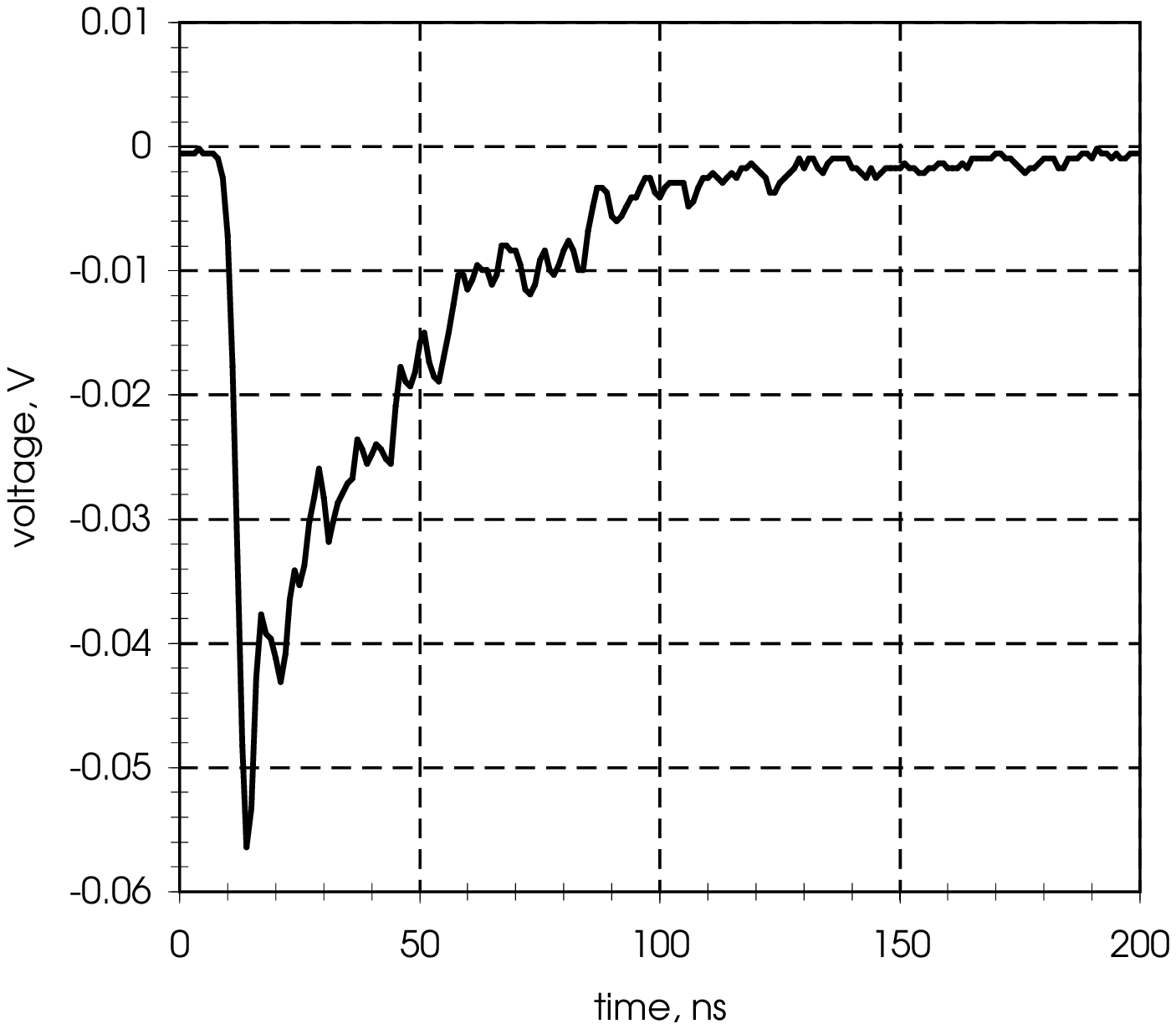,width=7.5cm}\epsfig{file=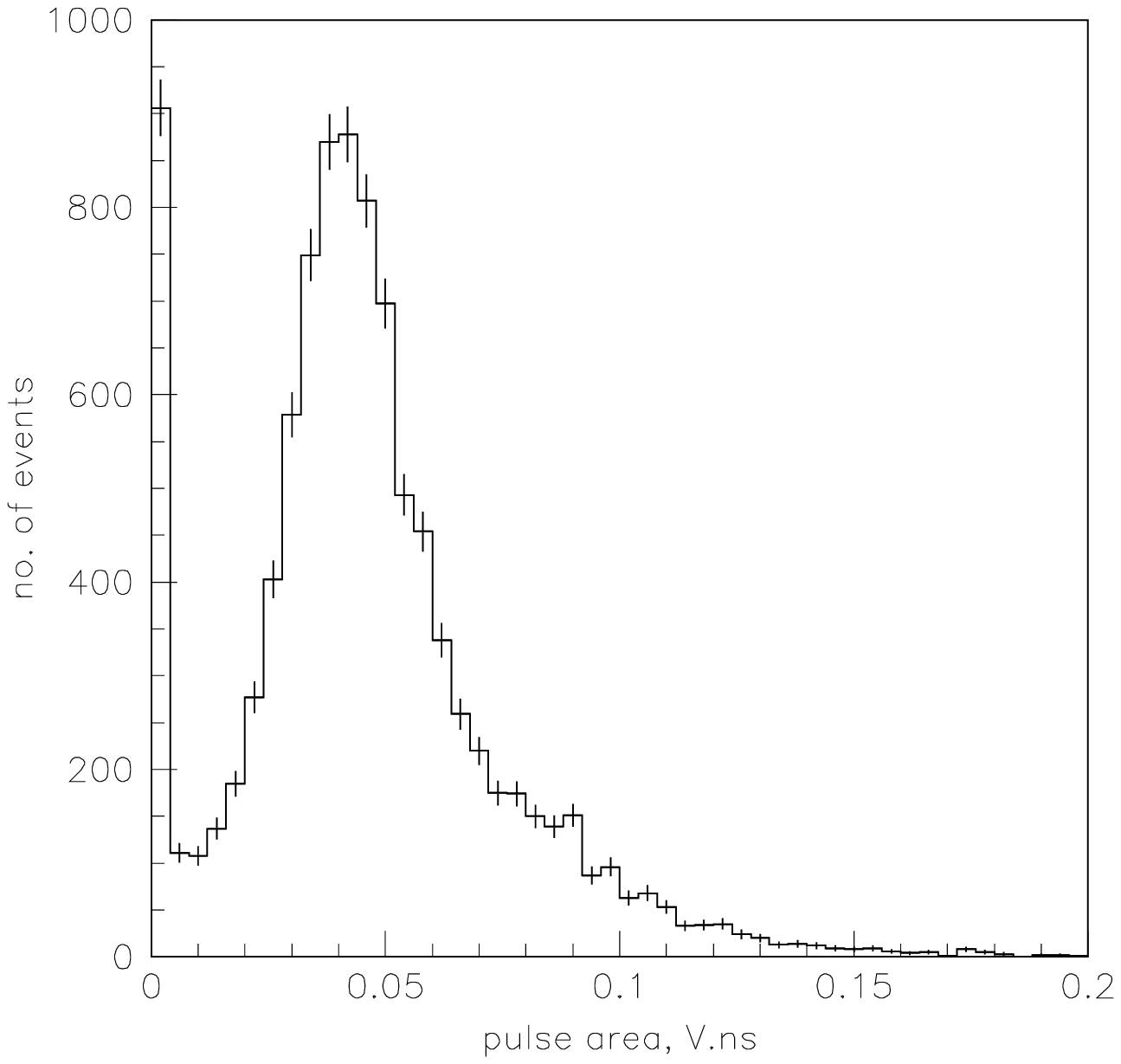,width=7.5cm
}}\caption{\small A typical primary scintillation pulse from a
59.6\,keV photon interaction and a single photoelectron spectrum
from one of the PMTs.  Both the spectrum and the scintillation
pulse were obtained with +2kV on the PMT anode; however there was
an additional x10 amplifier present for the spe measurement.}
\label{ScintillationPulse}
\end{figure}

Pulse height spectra, pulse waveforms and single photoelectron
spectra were collected from all PMTs during the first cold-run
both with the DAQ electronics just described and with a pulse
height analysis (PHA) set-up using a multichannel analyser (MCA).
These confirmed correct operation of all 31 PMTs in the array,
including $\sim$1000 crimped connections! A typical primary
scintillation pulse from a 59.6\,keV photon interaction is shown
in figure~\ref{ScintillationPulse}. This shows the characteristic
decay time of $\sim40$\,ns.  The single photoelectron spectra show
well resolved peaks and these were used to set the amplifier gains
in order to normalise all channels to the same overall gain.

\subsection{Scintillator performance}
LXe scintillates in the vacuum ultraviolet (VUV), near 175~nm,
with a yield comparable to the best scintillator crystals.  The
VUV luminescence is produced by the decay of singlet and triplet
states of the Xe$_2^*$ excimer. These can be formed directly by
excited atoms left by the interacting particle or as a result of
recombination into an excited state along the particle track
\cite{kubota79,hitachi83}.

\begin{figure}[ht]
\centerline{\epsfig{file=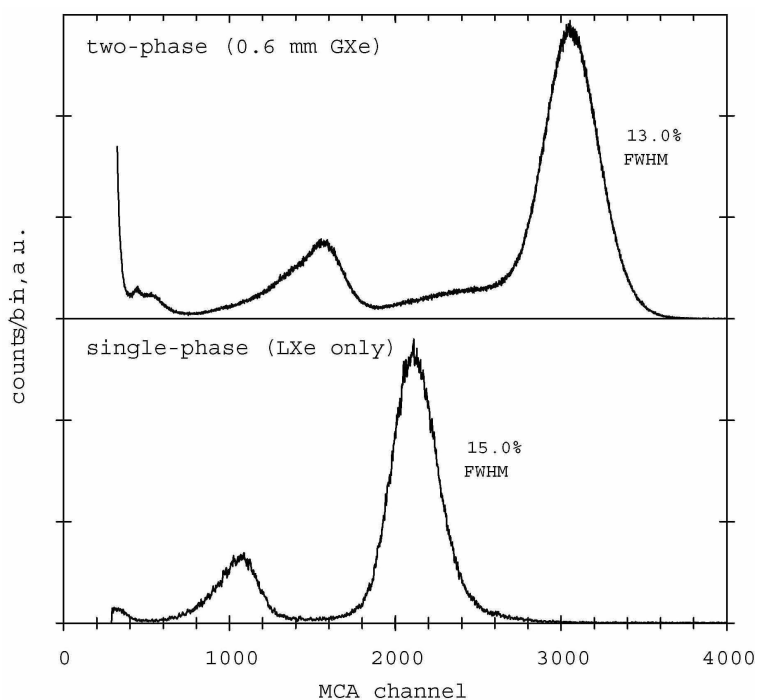,width=10cm}}
\caption{\small Pulse height spectra obtained from the $^{241}$Am
primary scintillation signals during the first cool-down test. The
two panels correspond to two different liquid xenon levels.  The
highest energy peaks in both spectra are at 59.6\,keV.}
\label{Am241spectra}
\end{figure}

Figure~\ref{Am241spectra} shows a typical spectrum taken from one
PMT when the internal $^{241}$Am sources were in place.  The plot
shows two spectra.  The bottom spectrum was taken with the whole
arrangement covered with liquid xenon.  The two spectral features
are the $59.6\,$keV line from $^{241}$Am and a blend of the
26.3~keV $^{241}$Am $\gamma$-ray with a 30~keV line resulting from
escape of Xe K-shell fluorescence photons.  Using the measured
single photoelectron spectrum from this PMT gives a signal level
of $\sim$12\,phe/keV. For this measurement there is no applied
electric field.  The top spectrum was taken with the liquid level
between the source and the PMT window.  The interaction occured in
the liquid phase and the improved light collection (up to
$\sim17$\,phe/keV is a result of total internal reflection at the
liquid gas interface due to the refractive index mismatch.  The
resolution from the two-phase spectrum is $\sim13\%$ FWHM.  More
detailed and extensive physics results from the first cold run
will be published separately\cite{chepel06}.

\subsection{Two-phase operation}
Once the radioactive sources used for the measurements in the
previous section had been removed the second and subsequent
cold-runs have successfully loaded the detector with liquid xenon.
A capacitive level-sensing system probes the liquid xenon height
with sub-mm accuracy at three locations in the chamber. A signal
from one of these coaxial capacitor structures, readout to 0.03~pF
accuracy, is shown in figure~\ref{stability}. In underground
operation, these sensors will be integrated with an active
levelling system in order to maintain the electrodes parallel to
the liquid surface, guarding the heavily shielded detector against
any structural deformation of the underground cavern.

\begin{figure}[ht]
\centerline{\epsfig{file=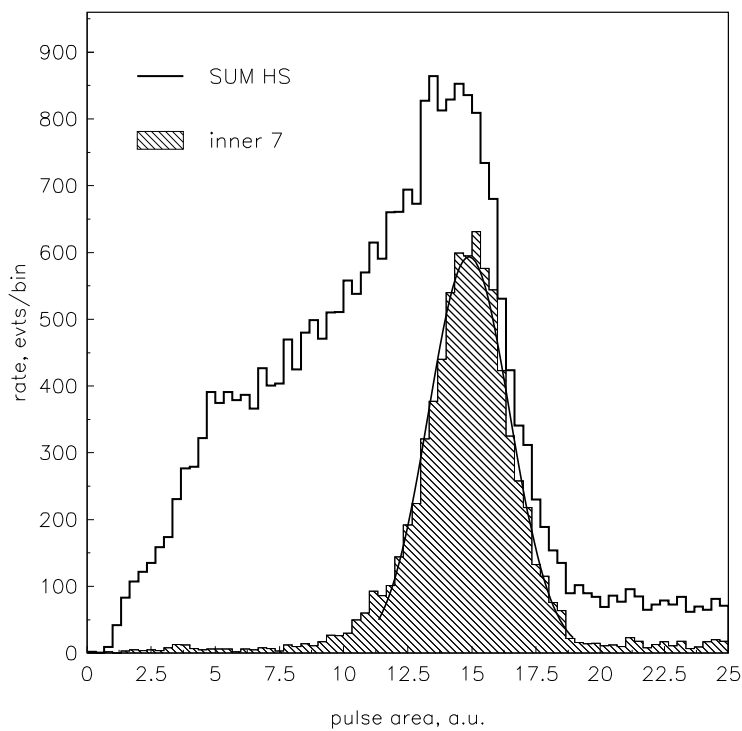,width=10cm}}
\caption{\small Pulse height spectra obtained from using an
external $^{57}$Co source above the instrument.} \label{co57}
\end{figure}

With the xenon filled to its nominal depth, but with no applied
electric field, figure~\ref{co57} shows $^{57}$Co spectra obtained
with an external uncollimated source placed above the detector.
Two spectra are shown, both reconstructed using the outputs from
all the PMTs. The shaded one, however, only includes events in
which the peak signal occurred in one of the inner seven PMTs.
This `collimated' spectrum has a FWHM of $\sim25$\%, which we
expect will improve once final corrections for PMT sensitivities
are done. The broad shoulder on the low side for the uncollimated
spectrum is purely due to light collection variation towards the
edge of the xenon volume (way outside the fiducial volume).

\begin{figure}[ht]
\centerline{\epsfig{file=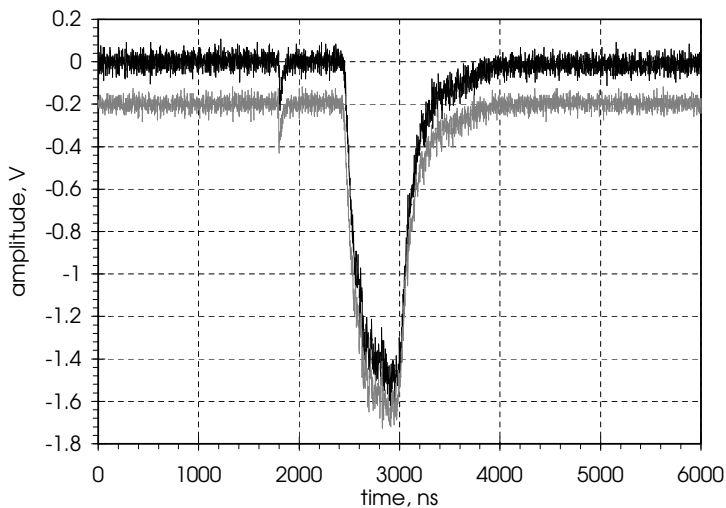,width=10cm}}
\caption{\small Summed waveforms from a $\gamma$-ray event showing
a fast primary pulse followed by the secondary wider pulse from
electroluminescence in the gas phase caused by ionisation drifted
from the event site.  The two traces shown are from the dual range
DAQ.  The low-sensitivity data have been multiplied by 10.}
\label{two_phase_single}
\end{figure}

After the zero-field tests, 13.5~kV were applied between the
cathode grid and the anode mirror, setting up a field of 3.0~kV/cm
in the liquid. Figure~\ref{two_phase_single} shows a typical
signal from a $\gamma-$ray interaction in the liquid. The fast
scintillation is the primary signal, S1, caused by direct
excitation created by the photoelectron. The second, broader
signal, S2, occurs when the ionisation also created has drifted to
the liquid surface and has been extracted into the gas phase. Once
in the gas phase the electric field is strong enough to cause
excitation leading to a burst of additional photons.  The time
delay depends on the depth at which the interaction happened and
the drift velocity at our operating fields is
$\sim$2.5\,mm/$\mu$s. The width of the secondary depends on the
gas gap and the electric-field in the gas.  The secondary emission
shows a flat plateau as the charges drift across the gap. The rise
and fall time is due to a combination of extraction dynamics,
diffusion and the gas scintillation time-constant. The ratio of
the two areas (S2/S1) is $\sim$150 as expected at this field.

\begin{figure}[ht]
\centerline{\epsfig{file=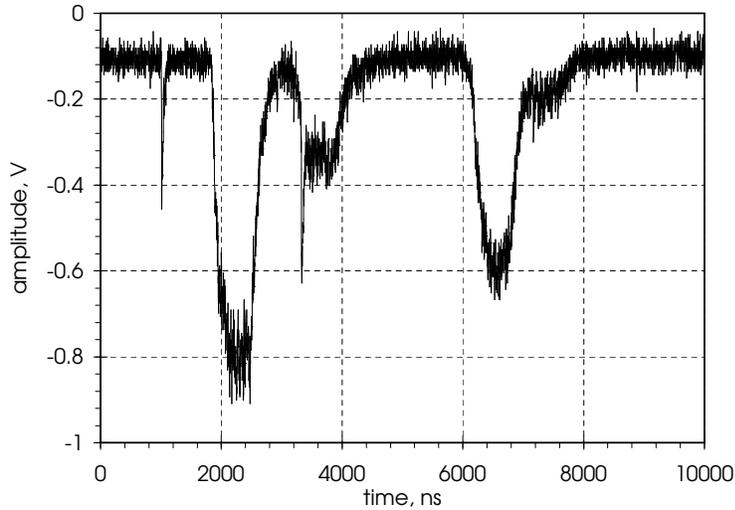,width=10cm}}
\caption{\small Summed waveforms from two overlapping $\gamma$-ray
events both showing fast primary signals followed by secondary
signals.} \label{two_phase_double}
\end{figure}

\begin{figure}[ht]
\centerline{\epsfig{file=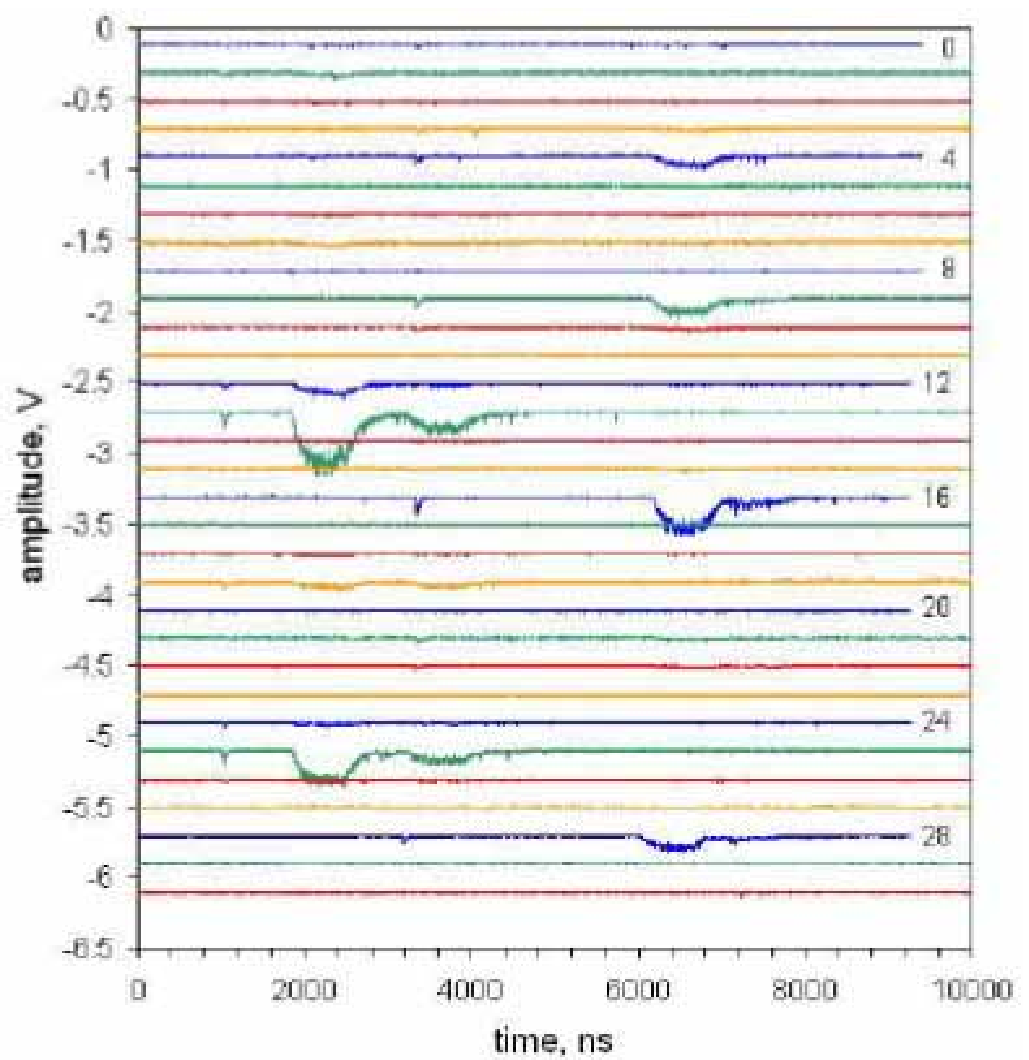,width=7cm}\epsfig{file=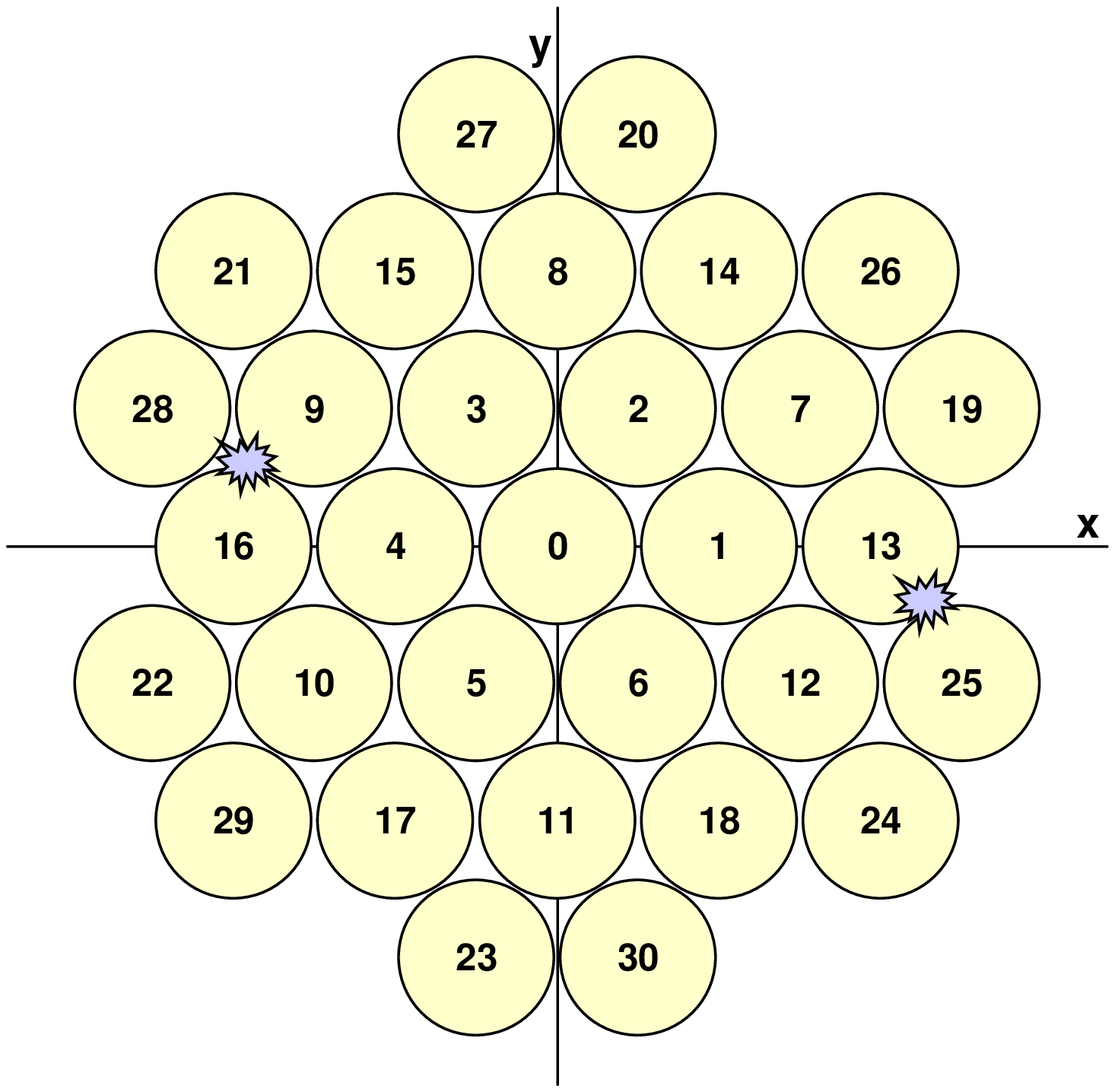,width=7cm}}
\caption{\small Individual waveforms from the 31 PMTs showing how
the two overlapping events can be separated using position
reconstruction.  The reconstructed positions are indicated on the
right-hand panel and the size of the symbols is representative of
the position resolution.} \label{multiple_sep}
\end{figure}

One of the key design drivers of ZEPLIN-III was the ability to
resolve each interaction point in the three dimensions. A position
reconstruction algorithm was developed from simulated datasets
which will provide sub-cm resolution in the horizontal plane
\cite{lindote05}. Even before this is applied to real data, this
spatial sensitivity is well demonstrated in figures
\ref{two_phase_double} and \ref{multiple_sep}, showing an event in
which two interactions have overlapped in time. Moreover there are
at least four secondary signals.  Without position sensitivity it
would not be possible to separate these two events just from the
summed signals.  However, looking at the individual PMT traces
(left-hand panel in figure~\ref{multiple_sep}) it is immediately
obvious that these two events have happened in very different
parts of the detector (right-hand panel) and they can be
unambiguously separated.  They are both double-Compton scatters.

\section{Summary}
The key design features of the ZEPLIN III instrument have been
described. The challenging and pioneering aspects of the
manufacturing technologies and procedures have been detailed and
first commissioning data have been presented to demonstrate the
successful completion of build of this instrument.  Further
extensive calibration work is now needed to prepare this
instrument for use as a dark matter detector.  To fully
charaterise ZEPLIN III much of this will need to be done
underground in a lower background environment.

%%%%%%%%%%%%%%%%%%%%%%%%%%%%%%%%%%%%%%%%%%%%%%%%%%%%%%%%%%%%%%%%%%%%%%%%%%%%%%
\section{Acknowledgements}

This work has been funded by the UK Particle Physics And Astronomy
Research Council (PPARC).  We would like to acknowledge the superb
copper machining achieved within the Imperial College Phyiscs
Department workshop led by R. Swain, and the development of new
welding techniques by The Welding Institute.

%%%%%%%%%%%%%%%%%%%%%%%%%%%%%%%%%%%%%%%%%%%%%%%%%%%%%%%%%%%%%%%%%%%%%%%%%%%%%%
%%%%%%%%%%%%%%%%%%%%%%%%%%%%%%%%%%%%%%%%%%%%%%%%%%%%%%%%%%%%%%%%%%%%%%%%%%%%%%

\end{document}